\documentclass[twocolumn]{emulateapj}

\newcommand{\boldnu}{\mbox{\boldmath{$\nu$}}}
\newcommand{\boldtheta}{\mbox{\boldmath{$\theta$}}}

\begin{document}

\title{Bayesian Inference from Observations of Solar-like Oscillations}
\author{B. J. Brewer$^1$, T. R. Bedding$^1$, H. Kjeldsen$^2$ and D. Stello$^1$}
\email{brewer@physics.usyd.edu.au}
\affil{$^1$Institute of Astronomy, School of Physics, A28, 
University of Sydney, NSW 2006, Australia \\ $^2$Department of Physics and Astronomy, University of Aarhus, Ny Munkegade, bygning 520, DK-8000 Aarhus C, Denmark}

\begin{abstract}
Stellar oscillations can provide a wealth of information about a star, which can be extracted from observed time series of the star's brightness or radial velocity. In this paper we address the question of how to extract as much information as possible from such a dataset. We have developed a Markov Chain Monte Carlo (MCMC) code that is able to infer the number of oscillation frequencies present in the signal and their values (with corresponding uncertainties), without having to fit the amplitudes and phases. Gaps in the data do not have any serious consequences for this method; in cases where severe aliasing exists, any ambiguity in the frequency determinations will be reflected in the results. It also allows us to infer parameters of the frequency pattern, such as the large separation $\Delta \nu$. We have previously applied this method to the star $\nu$ Indi \citep{nuindi}, and here we describe the method fully and apply it to simulated datasets, showing that the code is able to give correct results even when some of the model assumptions are violated. In particular, the non-sinusoidal nature of the individual oscillation modes due to stochastic excitation and damping has no major impact on the usefulness of our approach.
\end{abstract}


\section{Introduction}
Asteroseismology is fast becoming a standard tool in stellar astrophysics \citep[e.g.][]{2005JApA...26..123K}. The properties of a star's oscillations are determined by the star's structure and composition, so measurements of those oscillations have enabled astronomers to probe the interiors of stars. In particular, the frequencies of the various modes can provide information about the speed of sound at various depths in the star, and hence constrain its composition. Recently, there has been a rapid increase in the amount of data available for studying solar-like stars in this way \citep[for reviews, see][]{2003PASA...20..203B, 2003Ap&SS.284...21B, 2004soho...14..101K}.

The data analysis procedures used for asteroseismology have tended to be based on Fourier methods. For example, the first step taken by most authors analysing the raw time series data is usually the computation of the power spectrum, or periodogram. However, it has been shown \citep{bretthorst1} that the periodogram is only an optimal procedure if we are estimating the frequency of the oscillations with a model that assumes that only one frequency is present, or, if there are many frequencies, that they are well separated. In the case of solar-like oscillations, we expect the star to be oscillating in many modes simultaneously, some of which may be closely spaced. Furthermore, the modes are damped and stochastically excited, so each mode cannot be represented by a single sinusoid. There is also a guiding theory telling us that the frequencies of the modes are not arbitrary, but form a well defined pattern. This constitutes additional information that we might wish to take into account when analysing observations.

The desire to extract as much information as possible from an astronomical dataset, and to combine this with prior information from other datasets or theoretical considerations, has led to an explosion in the use of Bayesian methods \citep[e.g.][]{2005ApJ...631.1198G, 0047, cosmonest}. The basic format of Bayesian inference is as follows. Our state of knowledge about any set of parameters $\boldtheta$ is described by a probability distribution over the parameter space. With prior information and assumptions $I$, a prior distribution $p(\boldtheta|I)$ (read as the distribution for $\boldtheta$ {\it given} $I$) is assigned to describe what we know about the parameters before taking into account the current dataset. Then, the effect of taking into account data $\bold{D}$ is to modify our state of knowledge from the prior distribution $p(\boldtheta|I)$ to the posterior distribution $p(\boldtheta|\bold{D},I)$, given by

\begin{equation}\label{bayes}
p(\boldtheta|\bold{D},I) \propto p(\boldtheta|I)p(\bold{D}|\boldtheta,I).
\end{equation}

The distribution $p(\bold{D}|\boldtheta,I)$ is the probability distribution for the data given the parameters. When the data are fixed, this is a function of the parameters and is called the likelihood function. Hence, whether a model (a point in the parameter space) becomes plausible after taking into account data $\bold{D}$ depends on how plausible the model was before taking into account the data and on how well the model predicts the observed data. This kind of reasoning has led to the development of new methods of data analysis, as well as new interpretations and understanding of familiar methods\footnote{For example, a least squares fit of a model to some data can be interpreted as finding the most probable set of parameters, with a Gaussian distribution for the data points and no strong prior information.}. An example relevant to spectral analysis is the finding of \citet{bretthorst1} that the periodogram is proportional to the log of the posterior probability density for the frequency of an oscillating signal, under certain assumptions. For more details on Bayesian data analysis, including its relationship to common Fourier techniques, see the textbook by \citet{2005blda.book.....G}.

Typically, the datasets that are used for asteroseismology are time series of either the brightness (photometry) or the radial velocity of the stellar surface (from spectroscopy), measured at a discrete set of times $\{t_1,t_2,...,t_N\}$, and with each data point subject to an observational error $\{e_1,e_2,...,e_N\}$. This results in the observed values for the radial velocity (or magnitude) $y$ at each of the $N$ times:
\begin{equation}\label{noise}
y_i = f(t_i) + e_i
\end{equation}
where $f(t)$ is the true form of the signal, and the $\{e_i\}$ are the observational errors. If the star is oscillating in $M$ different (pure) modes, with frequencies $\{\nu_1, \nu_2, ..., \nu_M\}$, then the functional form of the signal is
\begin{equation}\label{signal}
f(t) = \sum_{j=1}^M\left (A_j cos(2\pi\nu_j t) + B_j sin(2\pi\nu_j t)\right)
\end{equation}
The goal of the data analysis is usually to use the observed data $\bold{y} = \{y_1, y_2, ..., y_N\}$ to infer the frequencies $\boldnu = \{\nu_1, \nu_2, ..., \nu_M\}$, the number of frequencies $M$ and possibly their amplitudes. However, the values of the frequencies are not arbitrary. For low degree p-mode oscillations, the frequencies are given approximately by
\begin{equation}\label{pattern}
\nu_{n,l} = \Delta \nu (n + \frac{1}{2}l + \epsilon) - l(l+1)D_0
\end{equation}
The values of the parameters of this pattern, $\Delta \nu$, $\epsilon$ and $D_0$, are also of interest as they are related to the sound speed at various depths in the star \citep{1994ApJ...427.1013B}. Throughout this paper, we will denote any parameters of the frequency pattern collectively by $\boldtheta$, and concentrate on finding these parameters and the frequencies. The amplitudes $\{A_i\}$, $\{B_i\}$ are regarded as nuisance parameters. In a recent paper \citep{nuindi}, we used a Bayesian method to estimate the large separation $\Delta \nu$ of the metal-poor subgiant $\nu$ Indi. The purpose of this paper is to describe the method in greater detail and demonstrate its effectiveness on simulated data.

\section{Method}
\subsection{Derivation}
As with all Bayesian calculations, the first step is to write down Bayes' theorem (Equation~\ref{bayes}) for the joint probability distribution for all of the unknown parameters (the frequencies $\boldnu$, the corresponding sine and cosine amplitudes $\bold{A}$ and $\bold{B}$, and the frequency pattern parameters $\boldtheta$), given the data $\bold{y}$ and the prior information or assumptions $I$:

\begin{equation}
p(M, \boldnu, \bold{A}, \bold{B}, \boldtheta | \bold{y}, I) \propto p(M, \boldnu, \bold{A}, \bold{B}, \boldtheta | I)p(\bold{y} | M, \boldnu, \bold{A}, \bold{B}, \boldtheta, I)
\end{equation}

\noindent where $p(M, \boldnu, \bold{A}, \bold{B}, \boldtheta | I)$ is the prior probability distribution for all of the unknown parameters, and $p(\bold{y} | M, \boldnu, \bold{A}, \bold{B}, \boldtheta, I)$ is the likelihood function, or the probability density for obtaining the observed data, given the parameters. From equations~\ref{noise} and~\ref{signal}, it is clear that knowing the $\theta$ parameters is irrelevant for predicting the observed data, if we already knew the frequencies and amplitudes. Therefore the explicit dependence on $\theta$ can be dropped from the likelihood function:

\begin{equation}
p(\bold{y} | M, \boldnu, \bold{A}, \bold{B}, \boldtheta, I) = p(\bold{y} | M, \boldnu, \bold{A}, \bold{B}, I)
\end{equation}

\noindent Expanding the prior probability density with the product rule, we have
\begin{eqnarray}
p(M, \boldnu, \bold{A}, \bold{B}, \boldtheta | \bold{y}, I) \propto p(M, \boldtheta | I)p(\boldnu, \bold{A}, \bold{B} | M, \boldtheta, I) \nonumber \\ \times p(\bold{y} | M, \boldnu, \bold{A}, \bold{B}, I) \\
\propto p(M | I)p(\boldtheta | I)p(\boldnu | M, \boldtheta, I)p(\bold{A}|M, I)p(\bold{B} | M,I) \nonumber \\ \times p(\bold{y} | M, \boldnu, \bold{A}, \bold{B}, I)
\end{eqnarray}
where we have made the assumptions that the prior probability densities for all of the parameters are independent (e.g. knowing a frequency would not tell us anything about the amplitude), and that knowing the $\theta$ parameters would tell us something about the frequencies, but nothing about the amplitudes.

\noindent Since we are not interested in the amplitudes, we can integrate out the $\bold{A}$ and $\bold{B}$ variables, to obtain the posterior distribution for M, $\boldnu$ and $\boldtheta$ alone:

\begin{eqnarray}\label{integral}
p(M, \boldnu, \boldtheta | \bold{y}, I) \propto p(M | I)p(\boldtheta | I)p(\boldnu | M, \boldtheta, I) \nonumber \\ \times \int p(\bold{A}|M, I)p(\bold{B} | M,I)p(\bold{y} | M, \boldnu, \bold{A}, \bold{B}, I) d^M \bold{A} d^M \bold{B}
\end{eqnarray}

\noindent If the observational errors $\{e_i\}$ are independent and have a Gaussian distribution with standard deviations $\{\sigma_i\}$ (the $\sigma$'s are usually given with the dataset, in the form of an estimated uncertainty for each data point), then the likelihood function is

\begin{equation}\label{likelihood}
p(\bold{y} | M, \boldnu, \bold{A}, \bold{B}, I) \propto e^{-\frac{1}{2}\sum_{i=1}^N\left(\frac{y_i - f(t_i)}{\sigma_i}\right)^2}
\end{equation}

\noindent We chose the prior distribution for the amplitudes $\{A_i,B_i\}$ to be independent Gaussians with mean zero and standard deviation $\delta$ (in applications, we chose $\delta$ = 3 ms$^{-1}$):

\begin{eqnarray}
p(\bold{A}|M, I) = \prod_{j=1}^M \frac{1}{\delta\sqrt{2\pi}} e^{-\frac{1}{2}(\frac{A_j}{\delta})^2} \propto \delta^{-M} e^{-\frac{1}{2\delta^2}\sum_{j=1}^M A_j^2}\\
p(\bold{B}|M, I) = \prod_{j=1}^M \frac{1}{\delta\sqrt{2\pi}} e^{-\frac{1}{2}(\frac{B_j}{\delta})^2} \propto \delta^{-M} e^{-\frac{1}{2\delta^2}\sum_{j=1}^M B_j^2}
\end{eqnarray}

\noindent With these choices, the integral in equation~\ref{integral} is a Gaussian integral which can be evaluated analytically \citep{bretthorst1}. We will write the final result simply as $L(M, \nu)$, since the result of evaluating the integral now plays the role of the likelihood function in a simpler inference problem in which the amplitudes no longer appear:

\begin{equation}\label{posterior}
p(M, \boldnu, \boldtheta | \bold{y}, I) \propto p(M | I)p(\boldtheta | I)p(\boldnu | M, \boldtheta, I) L(M, \nu)
\end{equation}

\noindent For the implementation of our code (described in section~\ref{mcmc}), a function was written that can calculate the likelihood $L$ for a given set of frequencies $\boldnu$. Another simplification that we will make is the assumption that the prior probability distributions for the frequencies, given the $\boldtheta$ parameters, are all independent, so that knowing one frequency would not provide any information about the other frequencies, if we knew the $\boldtheta$ parameters:

\begin{equation}
p(\boldnu | M, \boldtheta, I) = \prod_{j=1}^M p(\nu_i | M, \boldtheta, I) = \prod_{j=1}^M g(\nu_i;\boldtheta)
\end{equation}

\noindent where the form of the function $g$ will be specified in section~\ref{freqprior}; it is essentially the prior for the value of a single frequency, given the $\boldtheta$ parameters.

\subsection{Markov Chain Monte Carlo}\label{mcmc}
Since we are interested in what the data have to say about the $\boldtheta$ parameters, it would be useful if we could calculate the marginal distribution $p(\boldtheta|\bold{y},I)$. However, this is not achievable in practice. A more realistic approach is to use a Markov Chain Monte Carlo (MCMC) algorithm \citep{2005blda.book.....G} to generate samples of models from the posterior distribution in Equation~\ref{posterior}. Then, by looking only at the $\boldtheta$ values in the sample, we have effectively marginalised over all of the frequencies. In our code, we used a version of the Metropolis-Hastings algorithm \citep{nealreview}. Starting from a model with zero frequencies ($M=0$, $\boldnu = \emptyset$) and an arbitrary starting point for the $\boldtheta$ parameters, we updated the model by proposing a small change, and then accepting or rejecting with a certain acceptance probability. The acceptance probabilities for the different kinds of proposed changes were chosen so that the stationary distribution of the Markov Chain is the same as the target distribution of interest (the posterior distribution in equation~\ref{posterior}). These acceptance probabilities are shown in Table 1. If a proposed change to the model is rejected, the next model in the sequence is the same as the previous one. When this algorithm runs, the output of the code is a random sequence of models, each possibly slightly different from the last, where the diversity amongst the models is indicative of the uncertainty of any inference. To save memory, a subset of effectively independent models from this sequence may be used for any subsequent calculations. This is called ``thinning''. Our use of MCMC methods for detecting sinusoidal signals in a noisy time series is similar to those demonstrated by \citet{bretthorst2}, and also \citet{2005PhRvD..72b2001U}, in different contexts.
\begin{table}
\begin{center}
\caption{Proposal transitions and their acceptance probabilities}
\begin{tabular}{lcc}
\hline Proposal Transition & Acceptance Probability  \\
\hline Add a frequency \\ (chosen from the prior $g(\nu;\boldtheta)$) & $\min[1,\frac{L_{\rm new}}{L_{\rm old}}]$\\
\\ Remove a frequency & $\min[1,\frac{L_{\rm new}}{L_{\rm old}}]$ \\
\\ Shift a frequency \\ (symmetric proposal distribution) & $\min[1,\frac{L_{\rm new}g(\nu_{\rm new};\boldtheta)}{L_{\rm old}g(\nu_{\rm old};\boldtheta)}]$ \\
\\ Shift the $\boldtheta$ parameters \\ (symmetric proposal distribution) & $\min[1,\frac{p(\boldtheta_{\rm new}|I)\prod_{i=1}^M g(\nu_i;\boldtheta_{\rm new})}{p(\boldtheta_{\rm old}|I)\prod_{i=1}^M g(\nu_i;\boldtheta_{\rm old})}]$\\
\end{tabular}
\medskip\\
\end{center}
\end{table}

\subsection{Proposal Distributions}
As with all MCMC methods which use the Metropolis-Hastings algorithm, suitable probability distributions for the proposed transitions in Table 1 must be chosen, particularly for the transitions that shift a frequency or shift the $\boldtheta$ parameters. It is common to use a Gaussian or Normal distribution centred at the current value to propose the new value of a parameter, and the width of the proposal distribution is chosen to achieve a moderate acceptance rate of about 20-50 per cent. However, finding an appropriate width for the proposal distribution for each parameter is a time-consuming process. To avoid this, we simply chose all of the proposal distributions to be mixtures of 3 Gaussian distributions centred at the current value, with standard deviations covering several different orders of magnitude. Then, the ``best'' proposal distribution width is at least being used some of the time. If further optimization is required, many interesting tricks exist, for an example, see \citet{shortcut}.

\section{Simulated Data}\label{sim}
Before applying the algorithm to actual data, we applied it to simulations for which we knew the true values of the parameters. We aimed to see whether the algorithm successfully recovered the true values and whether the uncertainties returned by the program were reliable.

In order to keep the problem as simple as possible for the initial tests, we used a simplified version of the frequency pattern. We assumed that the frequencies were evenly spaced with spacing $\frac{1}{2} \Delta \nu$, starting from a central frequency $\nu_{\rm centre}$, such that the frequencies were
\begin{equation}\label{spaced}
\nu_i = \nu_{\rm centre} + i \frac{\Delta \nu}{2}, i=-5,-4,...,5
\end{equation}
As it is easier to infer a small number of parameters than a large number, we assumed that $\nu_{center}$ was equal to the frequency at which the power spectrum peaks, reducing the problem to that of estimating a single parameter, the spacing $\frac{1}{2} \Delta \nu$.

\begin{figure}
\begin{center}\includegraphics[width = 0.45\textwidth]{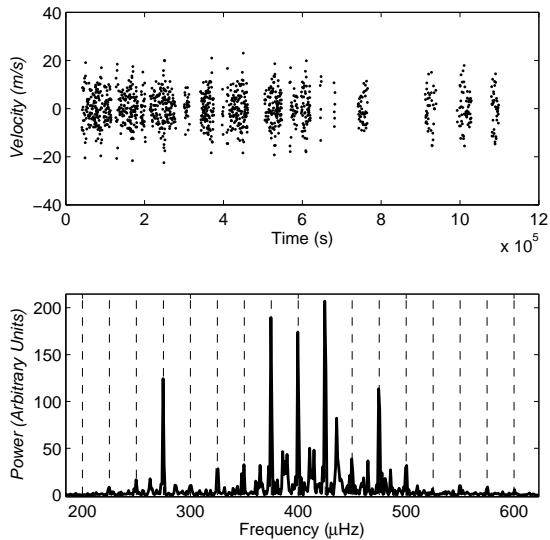}
\end{center}

\caption{The simulated dataset used for testing our method. The top panel shows the time series data (the error bars, of size 2 ms$^{-1}$, have not been plotted). The bottom panel shows the power spectrum of the time series, with several peaks visible at the input oscillation frequencies (shown with dotted lines).\label{simdata}}
\end{figure}

The time series was a sum of 17 sinusoids, with frequencies ranging from 200 $\mu$Hz to 600 $\mu$Hz, in steps of 25 $\mu$Hz. The phases were chosen at random, and the amplitudes were chosen at random from a normal distribution with a mean of zero and a standard deviation of 1 ms$^{-1}$. Gaps were introduced in the time series, since these are common in most astronomical time series datasets, due to the fact that observing can usually only be done at night, and interruptions may also occur due to poor weather\footnote{In fact, the observation times we used for the simulations were exactly the same as the actual observation times for $\nu$ Ind reported by \citet{nuindi}. There are 1201 data points in the time series.}. Finally, random Gaussian noise with a standard deviation of 2 ms$^{-1}$ was added to the data. This resulted in the time series shown in Figure~\ref{simdata}. The power spectrum is also shown, which has peaks at the input frequencies, but other alias peaks also exist.

Note that our model for the frequency pattern (Equation~\ref{spaced}) specifies that 11 frequencies are present, whereas our generated data set actually contains 17. This was done because we are unsure of the extent to which the frequency pattern holds, and to demonstrate that estimation of the frequency separation is robust to these assumptions; the only disadvantage of doing this is a slight decrease in the accuracy of the estimate of $\Delta\nu$ since we are only taking 11 out of the 17 frequencies into account.

\section{Choice of the prior $g(\nu;\boldtheta)$}\label{freqprior}
\begin{figure}
\begin{center}\includegraphics[width = 0.45\textwidth]{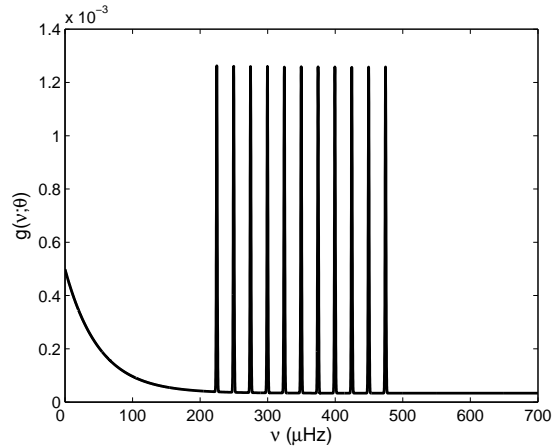}
\end{center}

\caption{An example of the prior $g(\nu;\boldtheta)$. There are three components: the low frequency exponential, the flat component, and the peaks where the putative relation of Equation~\ref{spaced} predicts that a frequency should be.\label{prior}}
\end{figure}

In this section, we consider the question of how to choose the prior for the value of a frequency, given the $\boldtheta$ parameters, in other words the function $g(\nu;\boldtheta)$. To choose an appropriate function, we must answer the question ``if we knew the values of the $\boldtheta$ parameters, what would this tell us about the value of a frequency''? Naively, we would expect the frequencies to fall exactly as predicted by equation~\ref{pattern}. In this case, our prior would be a sum of delta functions at the predicted values of the frequencies. However, we do not expect the relation to hold exactly, as it is an approximation. There may be slight deviations from it, and there may also be oscillation frequencies present which do not match the predicted frequency pattern at all, especially in an evolved star that exhibits mode bumping. Hence, we will need to use a slightly ``smeared out'' version of the prior, in which the delta functions are replaced by Gaussians, or some other suitable function. A uniform component was also included to catch any extraneous frequencies that do not fit in to our expectations at all. Also, we chose to include an additional exponential component, so low frequency noise (such as long term trends in the data) could be modelled. The final choice of the prior distribution was

\begin{eqnarray}
g(\nu;\boldtheta,\sigma) = \frac{1}{3\Lambda}e^{-\nu/\Lambda} + \frac{1}{3(\nu_{\rm max} - \nu_{\rm min})} \\ + \frac{1}{3N}\sum_{i=1}^{N}\frac{1}{\sigma\sqrt{2\pi}}e^{-\frac{1}{2}\left(\frac{\nu - f_i(\boldtheta)}{\sigma}\right)^2}
\end{eqnarray}

\noindent where $\Lambda$ is the scale length of the low frequency exponential (set to $\Lambda$=50 $\mu$Hz), ($\nu_{min}$,$\nu_{max}$) is the frequency range (we used a range from 0 to 700 $\mu$Hz), and $\sigma$ is a tolerance parameter, specifying how accurately we expect the predicted frequency pattern to hold; it sets the width of the Gaussian peaks. Note that this $\sigma$ is an additional variable and is not related to the noise $\sigma$'s in the likelihood function (Equation \ref{likelihood}). Since this is initially unknown, we included $\sigma$ as one of the parameters which we will infer, and assigned to it a uniform prior density over the positive real numbers. An example of the prior distribution $g(\nu;\boldtheta)$ for a typical value of the spacing and $\sigma$ is plotted in Figure~\ref{prior}.
It should be pointed out that our choice of the prior $g(\nu;\boldtheta)$ has been rather ad hoc. There is no reason why the three components should have equal weight (i.e. each component of the prior contains 1/3 of the probability), and a possible improvement to the method would involve allowing the weights to become free parameters as well, to be inferred from the data, as we have done with the $\sigma$ parameter\footnote{Preliminary tests indicate that this approach is viable and does not significantly increase the amount of CPU time required.}. Despite these assumptions, the code gives reasonable and useful results, which will be presented in Section~\ref{results}.

\subsection{Parallel Tempering}
The MCMC code as described above suffers from a serious problem. Early on in the run, frequencies tend to be added to the model in the locations that the initial values of the $\boldtheta$ parameters predict that they should be. Then, once these frequencies become established, the $\boldtheta$ values are unlikely to be changed by a large amount. In other words, the program can become stuck in a local minimum in the paramater space. Luckily, several techniques exist for overcoming these kinds of problems. Parallel tempering \citep{2005blda.book.....G} is one such method. Usually, several MCMC runs are run simultaneously, each with a progressively `softer' version of the likelihood function, achieved by raising the likelihood to some power that is less than 1 (the reciprocal of this power is called the temperature, and a higher temperature allows the model to move around more freely). Then, some proposed moves swap the models between chains with different temperatures. However, in this application, the problem does not arise because the likelihood function is multimodal, but rather because the prior $g(\nu;\boldtheta)$ is sharply peaked. Therefore, rather than softening the likelihood function for the high temperature chains, we softened the prior for the frequencies.

Define a tempered probability distribution $q_{T}(\boldtheta,M,\boldnu)$ (with temperature $T$) by

\begin{equation}
q_T(\boldtheta,M,\boldnu) \propto p(\boldtheta | I)p(M | I)  \left(\prod_{j=1}^M Z(T) g(\nu_i;\boldtheta)^{1/T}\right)  L(M, \boldnu)
\end{equation}

\noindent which is the same as the posterior (Equation~\ref{posterior}) but with a flattened version of the prior ($Z(T)$ is the normalisation constant for the prior). Then, proposed swaps between chains with models ($\boldtheta_1$,$M_1$,$\boldnu_1$) and ($\boldtheta_2$,$M_2$,$\boldnu_2$) and temperatures $T_1$ and $T_2$ are accepted with probability

\begin{equation}
\alpha = \min \left(1,  \frac{q_{T_2}(\boldtheta_1,M_1,\boldnu_1)q_{T_1}(\boldtheta_2,M_2,\boldnu_2)}{q_{T_1}(\boldtheta_1,M_1,\boldnu_1)q_{T_2}(\boldtheta_2,M_2,\boldnu_2)} \right).
\end{equation}

\noindent By using this acceptance probability, the set of Markov Chains samples from the distribution $q_{T_1}(\boldtheta_1,M_1,\boldnu_1)q_{T_2}(\boldtheta_2,M_2,\boldnu_2)...q_{T_N}(\boldtheta_N,M_N,\boldnu_N)$, where $T_1=1$. The output in the lowest temperature chain (with $T=1$) samples the posterior distribution (equation~\ref{posterior}), as required. We chose to run our simulations with 8 temperature levels, with a swap proposed every 10 steps. The temperatures used were $T_i = 1.2^{i-1}$ for $i=1,2,...,8$, as this allowed for an effectively flat prior at the highest temperature and an acceptance rate for the swaps of about 50 per cent. For different data sets, different tempering levels may be appropriate, but it is unlikely that any large modifications would be necessary. The performance of our algorithm is less sensitive to the choice of tempering levels than in conventional parallel tempering; this is because the inference is dominated by the likelihood rather than the prior, and we are only softening the prior. Hence, swaps of models between different temperature levels are often accepted.

\section{Results}\label{results}
\begin{figure}
\begin{center}\includegraphics[width = 0.45\textwidth]{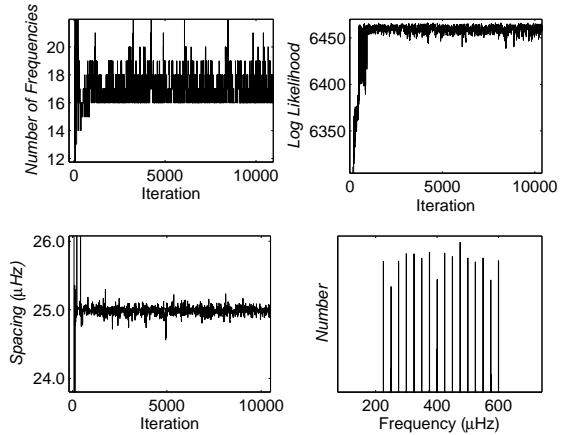}
\end{center}

\caption{Results from the Markov Chain Monte Carlo run, using the simulated dataset. The lowest frequency (200mHz) was not detected, as the amplitudes were generated randomly and this frequency must have had a very small amplitude.\label{box}}
\end{figure}

The output from the MCMC run is shown in Figure~\ref{box}. After an initial burn-in period, the distribution of the frequency spacing and the number of frequencies settles down to the posterior distribution. The inferred value of the frequency separation was (24.99$\pm$0.06)$\mu$Hz, a very accurate determination that is consistent with the true value of 25 $\mu$Hz. The histogram in the bottom right is an accumulation of the frequencies present in the models encountered throughout the run, and is very useful as a summary of the inference about what frequencies are present. By zooming in on these peaks, the uncertainty in any frequency can be easily measured, if this is of interest. The height of the peaks in this histogram are not particularly meaningful; the confidence that we have in the presence of a frequency is given by the fraction of the time that it occurs in the sequence of output models, whereas the height of these peaks in the histogram is probably more dependent on the binning that has been used. However, the total number of frequencies present within a short range of frequency space is directly proportional to the posterior probability that a frequency exists in this range, so a histogram with wider binning would have the heights proportional to the confidence of the frequency detections. In this case those probabilities are effectively 1. It is interesting to note that, while alias peaks exist in the power spectrum (Figure~\ref{simdata}), none are present in the output of the MCMC run. This is because the data are informative enough to conclusively decide which peaks are real and which aren't. This is typical behaviour, it is rare to find a genuine ambiguity. 

\begin{figure}
\begin{center}\includegraphics[width = 0.45\textwidth]{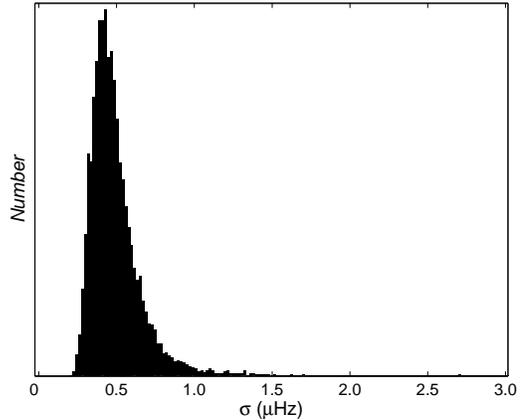}
\end{center}

\caption{The marginal posterior distribution for the tolerance parameter $\sigma$. The quality of the data is sufficient to provide evidence that the relation holds to a high degree of accuracy, so $\sigma$ is inferred to be quite small, and the peaks in the prior become quite sharp. As a result, the spacing between the frequencies is inferred very accurately.\label{sigma}}
\end{figure}

Interestingly, the number of frequencies has been inferred to be at least 16, but possibly up to about 20 (the true number was 17).  By examining the sample of models, it became clear that most of these additional frequencies were very close to the actual frequencies - in other words, it found that some of the frequencies were doublets. We believe this occurs because of our use of independent priors for each of the frequencies. For example, in reality, if we knew the frequency spacing and the values of a few of the frequencies, we would prefer a new frequency to be added in one of the gaps. However, since we have assumed independent priors for the frequencies, the program is just as likely to add another frequency into the model with the same value as an existing frequency. This is a minor flaw in our procedure, but it does not cause any large errors to be made, and it is worth keeping the independence assumption for its convenience. Of course, the data are also consistent with there being a few other frequencies of low amplitude, or there being very closely spaced frequencies, but the Bayesian ``Occam's Razor'' leads the program to suggest $M=$16 as the most probable solution.

Of course, this method may also be used when the predicted pattern of frequencies is not as simple as our evenly spaced example. With better data, it may be possible to infer many other parameters of this pattern, such as the small separation, and to search for departures from regularity.

\section{Effect of Stochastic Excitation and Damping}
\begin{figure}
\begin{center}\includegraphics[width = 0.45\textwidth]{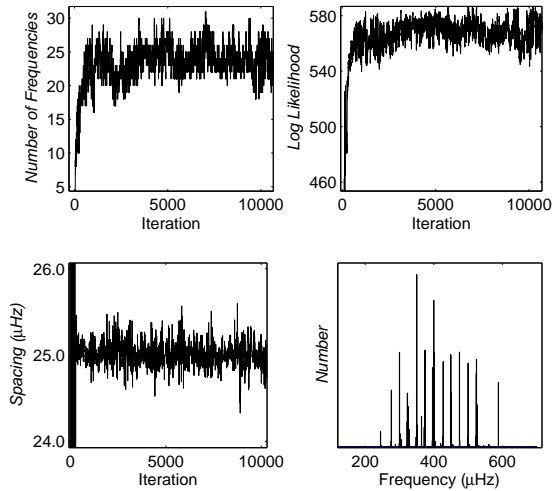}
\end{center}

\caption{Same as Figure~\ref{box}, but for a simulation in which each mode was stochastically excited and damped. More frequencies ($\sim$ 24) are required to explain the data, and the uncertainty in the value of the frequency spacing is increased.\label{box2}}
\end{figure}

\begin{figure}
\begin{center}\includegraphics[width = 0.45\textwidth]{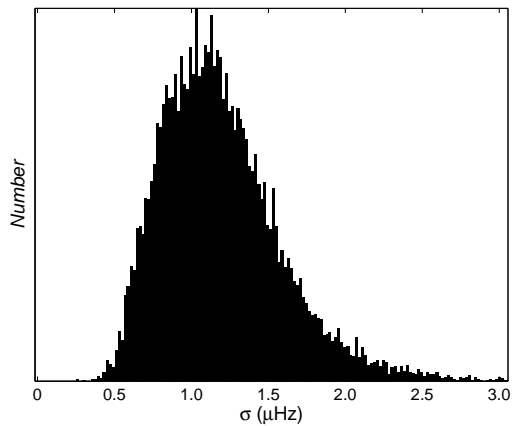}
\end{center}

\caption{Same as Fig~\ref{sigma} but for a simulation with stochastically excited and damped oscillations. The inferred value of $\sigma$ is much larger than in Figure~\ref{sigma}, and as a consequence the uncertainty about the frequency separation is increased.\label{sigma2}}
\end{figure}

The simulated dataset described in section~\ref{sim} is highly idealised. One important missing feature, which is an important aspect of solar-like oscillations, is the fact that the individual modes are not pure sinusoids. Instead, they are continually damped and stochastically excited by convection. This is one example of how the assumptions of our model may be violated by real data. It is important to test the method on some simulated data that includes this effect. To explore this, we produced a second simulated time series with the same frequency pattern, window function and noise level. We used the stochastic model described by \citet{Stello04}. The input amplitudes followed a gaussian distribution centred at 400 $\mu$Hz, with a FWHM = 200 $\mu$Hz and a maximum of 2 ms$^{-1}$. The mode lifetime was set to $3\,$days, in good agreement with measurements on the sun and other stars \citep{1997MNRAS.288..623C,2005ApJ...635.1281K,Stello05}. The results are displayed in Figure~\ref{box2}.
Since the modes are no longer pure sinusoids, some will need to be represented by more than one frequency. This explains the increase in the inferred number of frequencies, from 16 to about 24. In addition, the accuracy with which we can determine each frequency is reduced. As a result, the inferred value of $\sigma$ is larger (Figure~\ref{sigma2}), and the resulting uncertainty of $\Delta \nu$ is doubled, with the estimate of (25.01$\pm$0.13)$\mu$Hz for the spacing between the modes. It is encouraging that our method was still able to obtain the a correct result for $\Delta \nu$. As a check of the reliability of our method, we ran the code on five different realisations of stochastically excited and damped oscillations, and the inferred value of $\Delta \nu$ agreed with the true one (to within the quoted uncertainty) each time.
The fact that the inferred value of $\sigma$ is larger for the simulations with more damping (comparing Figures~\ref{sigma} and~\ref{sigma2}) suggests that we can measure the damping rate using this method. Previously, the damping rate has been measured by forward modelling to see the effect that damping and reexcitation has on the periodogram \citep{Stello05}.

\section{\'Echelle Diagrams}
\begin{figure}
\begin{center}
\includegraphics[width=3in]{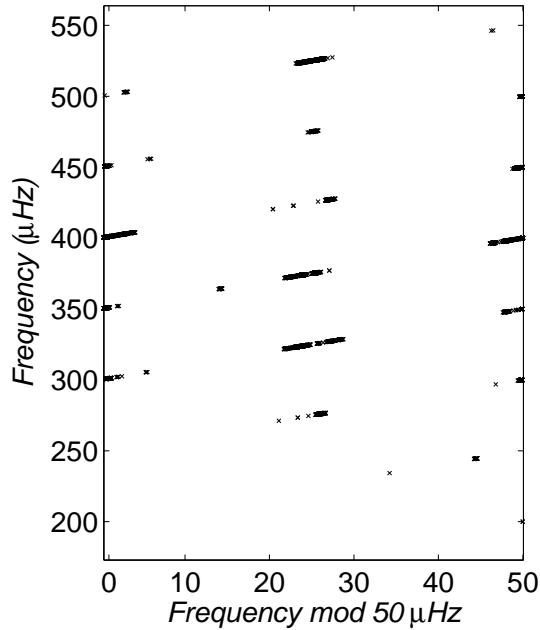}
\end{center}
\caption{\'Echelle diagram for the stochastically excited dataset. The large separation is 50 $\mu Hz$, which is twice the frequency spacing. The frequencies used for this plot are from the accumulated set of frequencies over the whole MCMC run, so each mode forms an almost horizontal line on this diagram, with the width of the line indicating the uncertainty about the frequency of that mode.\label{Echelle}}
\end{figure}

A commonly used tool in asteroseismology is the \'echelle diagram, a plot of frequency vs. frequency modulo the large separation $\Delta \nu$. The output of the MCMC program is a sequence of models for the star, allowing a sequence of \'echelle diagrams to be plotted, one for each model. The diversity of this sequence of plots would then indicate the uncertainty that we have about the true \'echelle diagram of the star. However, it is more convenient to have a single diagram, with the uncertainties in the frequencies plotted as error bars. To construct such a diagram, we used the accumulated frequencies from the MCMC run (i.e. the frequencies in the histogram at the bottom right of Figure~\ref{box2}). The resulting \'echelle diagram is displayed in Figure~\ref{Echelle}. Each frequency that was encountered is represented by a point in this diagram, and the spread for each mode arises due to the uncertainty in each frequency, and hence these lines can be interpreted as error bars.
\section{Conclusions}
In this paper, we presented a data analysis method for solar-like oscillation data, based on Bayesian inference and Markov Chain Monte Carlo methods. These methods are becoming very popular in astronomy and science in general, as they are often able to extract more information from the dataset than more ad-hoc approaches. In addition, reliable uncertainties in all inferred quantities are easily obtained.

The method we have presented here, which we have already applied to $\nu$ Ind \citep{nuindi}, has several advantages over more conventional approaches. For example, the CLEAN algorithm \citep{CLEAN}, which is based on iterative sine wave fitting, requires that the amplitude and phase be fitted as well. Then, when this fitted curve is subtracted, biases are introduced. This is because the fitted parameters are not exactly correct. Since our code fits all of the frequencies simultaneously, it does not suffer from this problem. Also, our method is able to take into account important prior information that we have about the expected pattern of frequencies, which other methods ignore. It could be argued that we aren't really sure that the prior information is correct, and therefore taking it into account may be giving us overconfident results. However, we have placed safeguards in our method to ensure that this does not happen - this was the purpose of introducing the $\sigma$ parameter and allowing it to be determined by the data. Importantly, we have also found that when some of the model assumptions are violated (in particular, if the modes are damped and stochastically excited), the method continues to give useful results.

Of course, this method is more computationally intensive than the usual power spectrum based methods\footnote{For our simulated data, usable output was obtained in a timescale of several hours on a modern PC.}. We would not recommend use of our method on very large datasets, for example those from the sun. In these cases, the power spectrum can be computed swiftly and is very informative. We believe that our method is best suited to those datasets where the data values (and hence the power spectrum) are noisy and incomplete, and aliasing causes difficulties in the interpretation of the power spectrum. As observational techniques improve, the number of target stars will increase and there will always be datasets for which this is the case. We hope that the approach we have presented here will prove useful in this field.

\acknowledgements
We acknowledge support from the Australian Research Council. BJB thanks Martin Hendry for helping to explain split-merge operations and allowing me to realise that I didn't need to use them, and also my PhD supervisor Geraint Lewis for giving me the freedom to explore different topics.

\label{lastpage}

\end{document}